\documentstyle[aps,prb,twocolumn,epsf,epsfig]{revtex}
\def\beq{\begin{equation}}
\def\eeq{\end{equation}}
\def\beqarr{\begin{eqnarray}}
\def\eeqarr{\end{eqnarray}}
\begin{document}
\draft
\twocolumn[\hsize\textwidth\columnwidth\hsize\csname @twocolumnfalse\endcsname

\title{Clustering in disordered ferromagnets:
The Curie temperature in diluted magnetic semiconductors}
\author{D. J. Priour, Jr. and S. Das Sarma}
\address{Condensed Matter Theory Center, Department of Physics,
University of Maryland, College Park, MD 20742-4111}

\date{\today}

\maketitle

\begin{abstract}
We theoretically investigate impurity correlation and magnetic 
clustering effects on the long-range ferromagnetic ordering in 
diluted magnetic semiconductors, such as 
$\textrm{Ga}_{1-x}\textrm{Mn}_{x}\textrm{As}$, using analytical 
arguments and direct Monte Carlo simulations.  We obtain an analytic
formula for the ferromagnetic transition temperature $T_{c}$ which 
becomes asymptotically exact in the strongly disordered, highly 
dilute (i.e. small $x$) regime.  We establish that impurity 
correlations have only small effects on $T_{c}$ with the 
neutrally correlated random disorder producing the nominally highest 
$T_{c}$.  We find that the ferromagnetic order is 
approached from the high temperature paramagnetic side through a 
random magnetic clustering phenomenon consistent with the  
percolation transition scenario.
\end{abstract}

]

\section{Motivation and Introduction}
The effect of disorder on magnetism is an old and important problem in 
condensed matter physics and materials science. 
Worldwide current activity~\cite{menosdiez,menosnueve,menosocho} in 
diluted magnetic semiconductors
(DMS) where cationic substitutional doping (by a few percent) of a 
semiconductor with magnetic impurities 
(e.g. $\textrm{Ga}_{1-x}\textrm{Mn}_{x}\textrm{As}$ with $x \approx 0.01-0.1$)
seemingly leads to an intrinsic ferromagnetic material with a relatively 
high Curie temperature ($T_{c} \approx 100K-200K$), has renewed 
vigorous recent interest in this important subject.  In particular,
the intrinsic mechanism of ferromagnetism (i.e. the physics underlying
the long-range ordering of the randomly distributed magnetic impurity 
moments) as well as the precise role of disorder (e.g. arising from 
the random distribution of magnetic Mn ions at the cation sites 
in the GaAs lattice) are substantial questions,
both from the fundamental perspective of
understanding the competition between disorder and magnetic 
interactions as well as the technological perspective of the 
fledgling subject of `spintronics' (or spin electronics) where the 
projected seamless integration of magnetics and electronics on a single 
ferromagnetic semiconductor chip could lead to a new paradigm in 
microelectronics~\cite{menossiete}.  In this article, we study theoretically 
the competition between quenched disorder (i.e. the random spatial distribution
of the magnetic impurities) and magnetic correlations in DMS materials 
using analytical arguments supported by extensive large-scale Monte Carlo
simulations on a disordered Heisenberg model.  Our main findings are that (1)
site disorder by itself has a relatively small effect on the magnitude of 
the DMS ferromagnetic transition temperature $T_{c}$,
and (2) global ferromagnetic ordering in DMS is approached from the high-temperature 
(i.e. $T > T_{c}$) paramagnetic phase through the formation of random 
disconnected magnetic clusters of increasing size as 
$T$ approaches $T_{c}$ from above, which coalesce at $T = T_{c}$ leading to a magnetic
percolation transition, similar to what was postulated in ref.~\cite{menosseis}.  
We believe that our 
theoretical findings are quite general, and 
should apply to all diluted site disordered ferromagnetic
materials including doped magnetic 
oxides~\cite{menoscinco,menoscincopuntocinco}, 
which are insulators or poor metals (with short mean free paths comparable to 
lattice spacings).
In particular, we speculate that the results presented herein are sufficiently general
to be valid for {\it all} site disordered ferromagnets (with localized impurities 
providing the underlying magnetic moments) in the dilute (or, equivalently, strongly 
disordered) regime, though we discuss our work entirely in the
context of $\textrm{Ga}_{1-x}\textrm{Mn}_{x}\textrm{As}$ since this is by far the 
most extensively studied DMS material experimentally and theoretically.  

A key 
feature~\cite{menosdiez,menosnueve,menosocho,menossiete,menosseis,menoscinco,menoscincopuntocinco} of
DMS materials is that the 
ferromagnetic interaction between the impurity magnetic moments is an 
indirect exchange interaction mediated by the semiconductor carriers (which are holes in 
$\textrm{Ga}_{1-x}\textrm{Mn}_{x}\textrm{As}$).
The `standard' model for DMS 
ferromagnetism was developed~\cite{menoscuatropuntocinco} more than forty years ago,
and has recently been 
rediscovered~\cite{menoscuatropuntosetentaycinco}.
This model, which we refer to as the VCA model, is a simple 
mean-field model in both the spatial disorder (i.e. a continuum virtual 
crystal approximation, VCA) and the thermal fluctuations (i.e. a Curie-Weiss
mean field theory for the ferromagnetic ordering of the impurity moments) 
with the magnetic interaction between the impurity moments being mediated
by the indirect exchange interaction arising from the carrier-induced 
RKKY-Zener mechanism~\cite{menosdiez,menosnueve,menosocho}.  
Although the extreme simplicity of the VCA 
model, for example its analytic prediction of $T_{c} \propto mJ_{pd}^{2}n_{c}^{1/3}x$
(where $J_{pd}$ is the effective p-d exchange interaction strength between 
the p-type holes in GaAs and the Mn d-level, and $m$, $n_{c}$ are the effective 
hole mass and density, respectively), has led to its widespread application in the 
DMS literature, we emphasize here that the model is conceptually 
wrong since it predicts a monotonically increasing ferromagnetic 
transition temperature with increasing carrier density, impurity density,
and magnetic coupling strength - in fact, all three trends are conceptually
incorrect~\cite{Chat} (and the simple VCA model provides no mechanism for estimating its 
regime of validity in the $J_{pd}$, $n_{c}$, $x$, $T$, and disorder parameter 
space).  The questions (e.g. impurity correlation, magnetic 
clustering) being asked in this work are simply beyond the scope of VCA 
due to its complete neglect of site disorder and thermal fluctuations.
We also find that in the strong disorder limit (i.e. highly diluted case of 
$x \ll 1$), where the VCA fails most miserably, $T_{c}(J_{pd},x,n_{c})$ has a 
{\it qualitatively} different behavior than that predicted by the standard DMS 
VCA mean field theory.

\section{Aspects of the Calculation}
     We assume Mn ions randomly occupy only Ga sites in the GaAs 
     zinc-blende lattice (fcc) with a lattice constant $a$.
We operate in the 
(weakly) metallic limit and assume the carrier-mediated effective 
Mn-Mn indirect exchange interaction to be of the RKKY form.  However,  
with DMS systems being at best poor or bad metals with a mean free path 
typically of the order of 1-2\AA which is less than $a$, 
it is important to include the 
effects of a finite carrier mean free path; we do this by 
introducing a cutoff $l (\lesssim a)$ in the range of the RKKY function.  The large
value ($S=5/2$) of the impurity moment spins allows a treatment of the 
Mn spins as classical Heisenberg spins.
Hence, our Hamiltonian is given by
\begin{equation}
{\mathcal H} = \sum_{ij}J(r_{ij}){\bf S}_{i} \cdot {\bf S}_{j},
\label{eq:eq1}
\end{equation}
where $r_{ij}$ is the separation between moments $i$ and $j$, and $J(r)$ is 
the damped RKKY range function given in three dimensions by 
\begin{equation}
J(r) = J_{0}e^{-r/l}r^{-4}[\sin(2k_{\mathrm{F}}r)-2k_{\mathrm{F}}r\cos(2k_{\mathrm{F}}r)];
\label{eq:eq2}
\end{equation}
$k_{\mathrm{F}} = (\frac{3}{2} \pi^{2} n_{c})^{1/3}$ is the Fermi wave number and
$n_{c}$ is the hole density. 
$J_{0}(>0)$ is related to the local Zener coupling $J_{pd}$ between the
Mn local moments and hole spins, $J_{0} \propto mJ_{pd}^{2}$ where
$m$ is the hole effective mass (for
convenience, spin units are used with the factor $S^{2} = (5/2)^{2}$ absorbed
into $J_{0}$).  While $l$ sets the range of interaction between moments,
$l_{s} \equiv n_{i}^{-1/d}$ (where $n_{i}$ is the volume Mn density) gives
the mean inter-impurity separation and provides a measure of the disorder
strength with $l_{s} \approx a$ for weak disorder and $l_{s} \gg a$ for 
strong disorder.  Note that by definition the strong disorder limit
is equivalent to the dilute limit, which is the situation appropriate for 
DMS materials.  For weak disorder with a long range interaction 
(i.e for $l \gg \left \{ l_{s},a \right \}$), Curie Wiess Mean Field Theory (MFT)
fares well (though lattice MFT~\cite{menoscuatropuntoocho} 
is an important refinement
in addressing the 
discrete crystal structure); we show here that strongly disordered ferromagnets 
can also be viewed in terms of a limiting theory in which the Curie temperature 
asymptotically approaches 
\begin{equation}
T_{c} = \frac{\eta}{k_{B}}J(2r_{c}n_{i}^{-1/d}), 
\label{eq:eq3}
\end{equation}
where
$\eta$ is a constant of order unity and $r_{c}$  
is the dimensionless critical radius for
random sphere percolation with $r_{c} = .4436$~\cite{tres} (We concentrate 
entirely on the currently experimentally relevant three-dimensional $(d=3)$
DMS systems in this paper).
It is clear that this $T_{c}$
expression, $T_{c} \sim J(n_{i}^{-1/d})$, bears little resemblance to the VCA
result $T_{c} \propto n_{i} \int J(r) d^{d}r$.
   
   To calculate quantities such as $T_{c}$ and the spontaneous 
magnetization $\bf{M}$, we use a hybrid Monte Carlo technique combining 
Wolff Cluster~\cite{cinco} and Heat Bath~\cite{seis} moves.  The Wolff 
moves overcome critical slowing down near $T_{c}$, and Heat Bath moves  
ensure that ergodicity is attained efficiently by thermalizing 
spins weakly coupled to their neighbors.   
We calculate $T_{c}$ by locating the 
crossing temperature of Binder cumulants $U$~\cite{ocho} for two different 
system sizes; in our case,   
$U = 1 - (1/3)[ \langle \left| {\bf M} \right|^{4} \rangle ] / [ \langle \left|
\bf{M} \right|^{2} \rangle ]^{2}$, where angular and square brackets denote
thermal and disorder averaging, respectively.  
Since we invariably have $\langle N_{\mathrm{imp}} \rangle \equiv n_{i}L^{3} \ge 2 \times 
10^{4}$, we have found it useful to 
exploit our large system sizes; calculating magnetization moments
${\left| {\bf M} \right|^{i}}$ by simply raising the total 
magnetization ${\bf M}$ to the $i$th 
power disposes of information that can be gleaned by dividing 
the system into $n_{s}$ subsystems and averaging 
${\left| {\bf M} \right|^{i}}$ evaluated for each of these.
($n_{s} = \langle N_{\mathrm{imp}} 
\rangle^{1/2}$ has proven to be a reasonable choice).  Operating in 
this manner, we use 
as few as 50 total Monte Carlo sweeps ($n_{\mathrm{eq}}=25$ for equilibration and 
$n_{\mathrm{av}}=25$ for statistical sampling) with no compromise in accuracy.   
We have checked for each 
calculation the adequacy of the equilibration stage by comparing 
``cold starts'' (systems prepared with perfect ferromagnetic order) and 
``hot starts'' (systems with initially randomized spins), finding in 
each case agreement to within random statistical error (confined to 
less than $1\%$ for all results reported here).

\section{Correlated Magnetic Moments}
\begin{figure}
\centerline{\psfig{figure=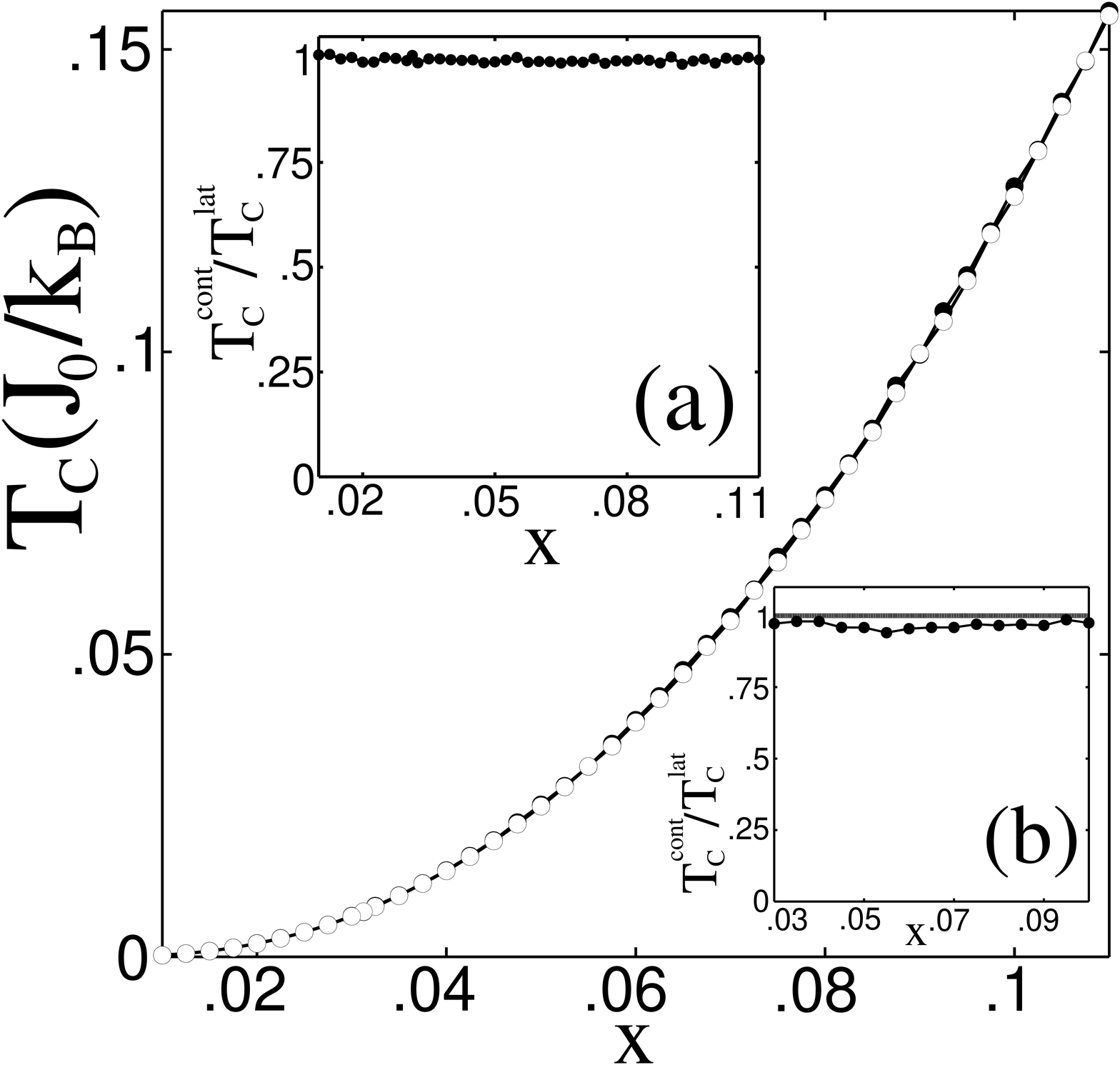,width=3.2in}}
\centerline{\psfig{figure=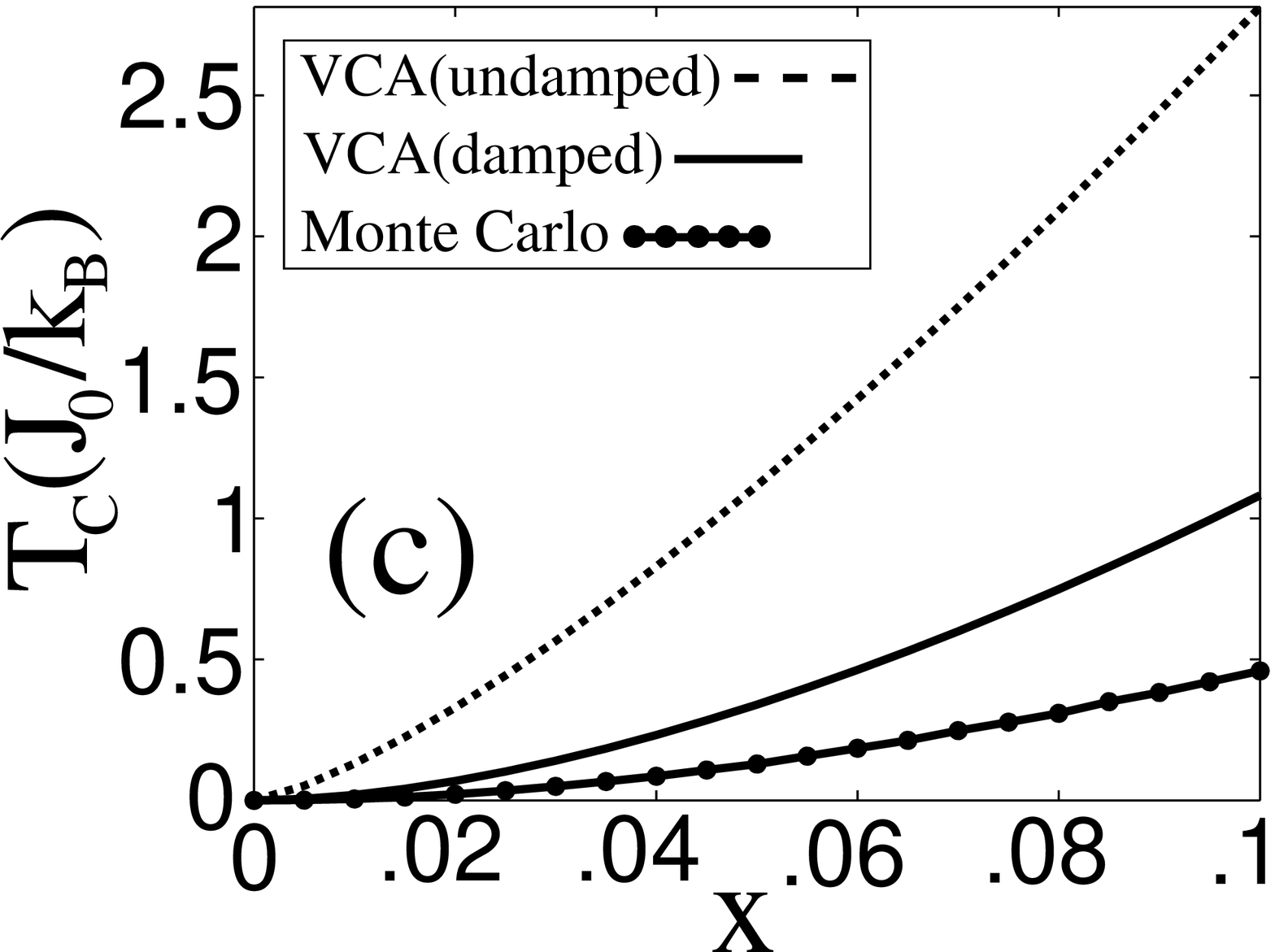,width=2.5in}}
\caption{In the main graph, $T_{c}^{\mathrm{cont}}$(continuum) and 
$T_{c}^{\mathrm{lat}}$(discrete)
are graphed with continuum $T_{c}$'s shown 
as closed circles and discrete $T_{c}$'s plotted as open circles;
$n_{c}/n_{i} = 0.1$ and $l/a = 0.5$.  In inset (a), ratios 
$T_{c}^{\mathrm{cont}}/T_{c}^{\mathrm{lat}}$ are 
shown; again, $n_{c}/n_{i} = 0.1$ and $l/a = 0.5$.  The graph 
in inset (b) shows $T_{c}^{\mathrm{cont}}/T_{c}^{\mathrm{lat}}$ for  
$n_{c}/n_{i} = 0.1$ and $l/a = 0.25$ with the horizontal dark reference
line indicating unity.  Panel (c) shows VCA (solid and broken lines for 
damped and undamped cases, respectively) and  
Monte Carlo $T_{c}$'s  on the same graph with 
$n_{c}/n_{i} = 0.1$ and $l/a = 1.0$ (the damping refers to the 
inclusion of mean free path effects).} 
\label{Fig:fig2}
\end{figure}

Relaxing the
requirement that magnetic impurities reside on the discrete fcc lattice in
favor of continuously distributed Mn is a simple way to examine the
sensitivity of the ferromagnetic state to how the Mn ions are distributed.
In principle, the discrete and continuum cases represent very different 
situations. 
Even so, it its evident in Figure~\ref{Fig:fig2}
that Curie temperatures for the discrete ($T_{c}^{\mathrm{lat}}$)
and continuum ($T_{c}^{\mathrm{cont}}$) cases (shown for 
more than a decade of $x$ values) are in close agreement.  The main
graph shows critical temperatures directly for $l/a = 0.5$, inset (a) depicts 
the ratio of continuum and lattice $T_{c}$'s for $l/a = 0.5$, and inset (b) displays the 
same result for the even smaller cutoff $l/a = 0.25$. 
In the continuum model, it is possible for magnetic impurities to cluster 
more tightly than in the discrete case, and the spins in such clusters
are therefore more strongly aligned.  However, the enhanced intra-cluster
coupling of the compact clusters tends to be offset by a weaker coupling to 
neighboring spins (since the overall constancy of $x$ requires that 
compact structures be more isolated, on average, than more diffuse Mn
assemblies).  The close agreement of $T_{c}^{\mathrm{cont}}$ and
$T_{c}^{\mathrm{lat}}$ in Figure~\ref{Fig:fig2} suggests that these two effects 
nearly precisely cancel.  It is important to note that the close
similarity of continuum and discrete $T_{c}$'s does not justify the 
VCA result~\cite{menoscuatropuntosetentaycinco}.  
The latter, given by 
\begin{equation}
T_{c}^{\mathrm{VCA}}=
4\pi \frac{J_{0}}{3}x \int r^{2} J(r)dr
\label{eq:eq4}
\end{equation}
does not take into account 
thermal and disorder fluctuations
(important in the dilute DMS regime).  The disparity between the VCA
and Monte Carlo results is highlighted in panel (c) of Figure~\ref{Fig:fig2},
and is even greater if the damping factor $e^{-r/l}$ is not 
taken into account.
\subsection{impurity clustering}
\begin{figure}
\centerline{\psfig{figure=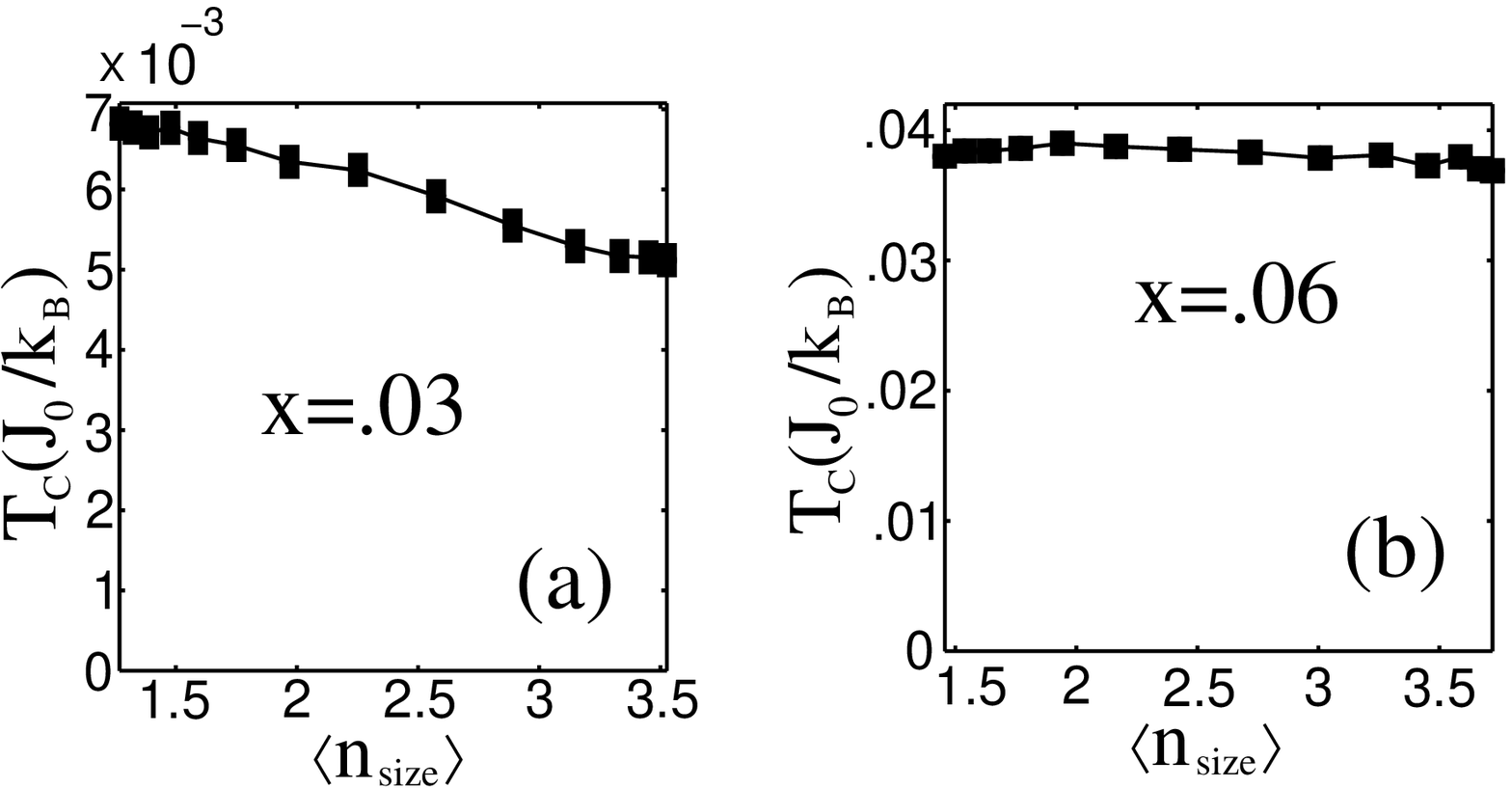,width=3.2in}}
\centerline{\psfig{figure=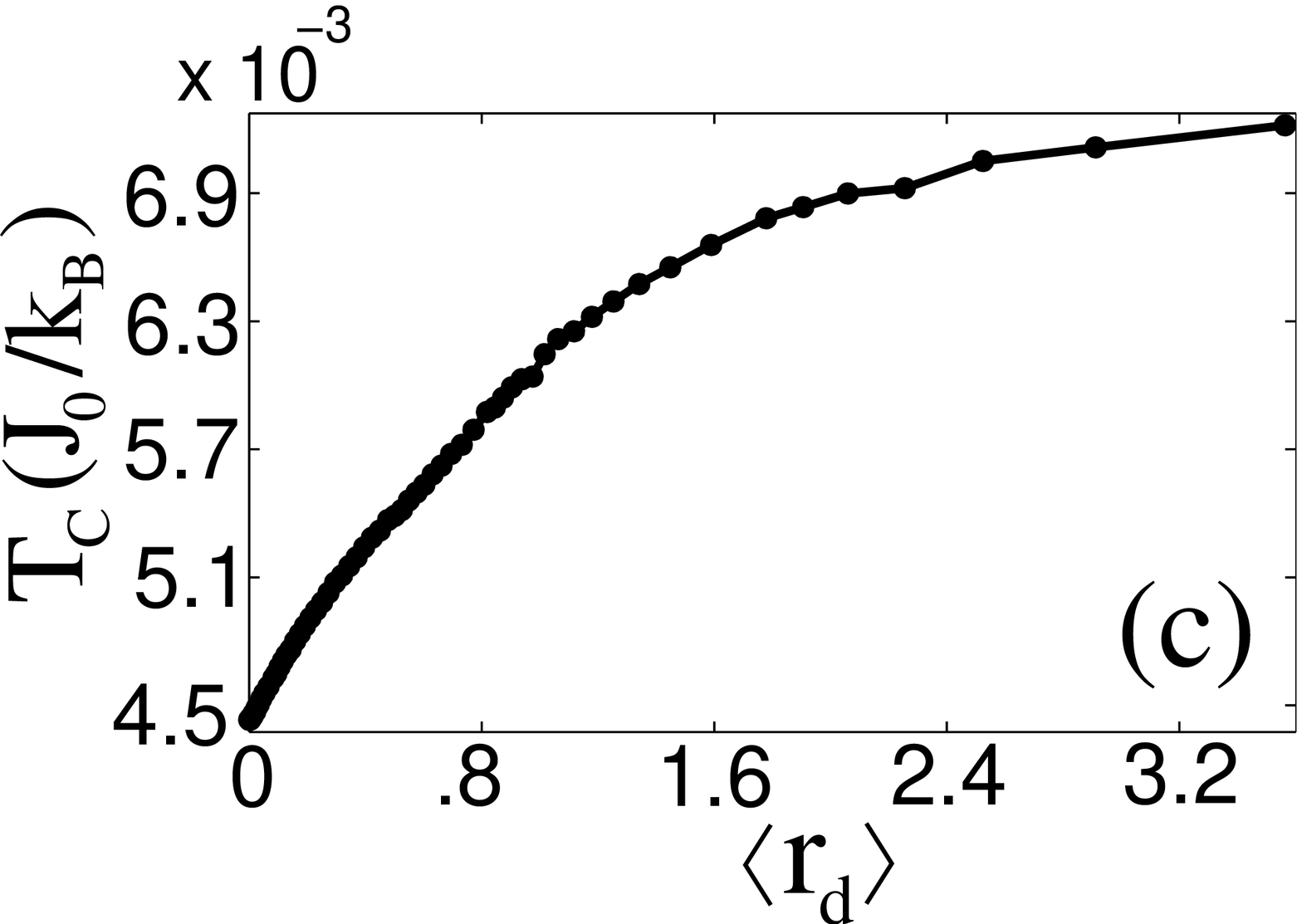,width=2.25in}}
\caption{Curie Temperatures for impurity clusters (panels (a) and (b)) and 
impurity superlattices (panel (c)). 
In (a) and (b), where $x$ is respectively $3 \%$ and $6 \%$, 
$T_{c}$ is plotted versus $\langle n_{\mathrm{size}} \rangle$, 
the mean cluster size.   
In (c), $T_{c}$ is graphed versus $\langle r_{d} \rangle$, the mean 
deviation from superlattice sites.  For (a), (b), and (c), $n_{c}/n_{i} = 0.1$
and $l/a = 0.5$.}
\label{Fig:fig3}
\end{figure}

     To examine impurity correlations
in a way that is continuously tunable,
we introduce a local attractive interaction between 
impurity moments yielding the configurational energy  
$E_{\mathrm{conf}} = -E_{0}\sum_{nn}(\mathrm{bonds})$ where 
$\sum_{nn}(\mathrm{bonds})$ is the total number of bonds 
shared by neighboring Mn ions.  We prevent the formation of a single 
large cluster by limiting the lateral dimension of clusters, 
requiring in this case that the size along any axis be smaller than $a$.  
We vary the extent of impurity clustering via an annealing temperature $T_{A}$;
large $T_{A}$ values lead to weaker clustering, while in the low 
$T_{A}$ limit, impurities tend to be more exclusively bound in 
the complexes mentioned above.
Configurations are prepared by subjecting initially neutrally 
correlated samples to an equilibration stage at $T_{A}$ (consisting
of 500 Metropolis sweeps), and the extent of clustering 
in the resulting sample is characterized by $\langle n_{\mathrm{size}} \rangle$,
the mean cluster size.  Impurity moments locked in clusters tend to 
align due to their proximity, but the large 
inter-cluster spacings makes the alignment of cluster moments with 
each other more readily disrupted by thermal fluctuations.  In fact as  
$x$ is decreased, 
one ultimately obtains an essentially paramagnetic state composed of large
but very weakly coupled cluster moments.  In Fig.~\ref{Fig:fig3}(a)
and Fig.~\ref{Fig:fig3}(b) the impact of clustering on $T_{c}$ is shown 
for two different Mn concentrations $x$ for clustering ranging from very 
weak to quite strong.
For $x = 0.03$ (the case depicted in Fig.~\ref{Fig:fig3}(a)), 
the overall Mn density is sufficiently low that one does 
see a slight dimunition of $T_{c}$ for increasing clustering.  
However, when $x$ is doubled to 
$x = 0.06$ (shown in Fig.~\ref{Fig:fig3}(b)), the $T_{c}$ curve 
is essentially flat.  Hence, even for fairly small Mn densities,
the effects of enhanced intra-cluster alignment 
and diminished inter-cluster coupling are roughly the same in magnitude,
leading to a weak net effect.  Thus impurity correlation or clustering does not seem 
to have much effect on $T_{c}$.  We mention as an important caveat
that in $\textrm{Ga}_{1-x}\textrm{Mn}_{x}\textrm{As}$, two closely placed 
Mn ions (e.g. a substitutional and interstitial Mn atom in the same unit cell) 
experience very strong nearest-neighbor short-range {\it direct 
antiferromagnetic} exchange coupling.  This would lead to~\cite{menoscuatropuntoocho} 
a strong suppression of ferromagnetic $T_{c}$ at large values of $x$ if 
there is substantial Mn clustering.  This effect is not included in our 
model.  

\subsection{impurity superlattices}
In a sense opposite to inducing impurity clustering is 
arranging Mn ions in superlattices, which tends to break up impurity 
clumps to form a homogeneous lattice.    
Again, we begin
with perfect superlattices and 
introduce deviations by annealing at the ``disordering temperature'' 
$T_{\mathrm{dis}}$; the mean displacement of impurities 
$\langle r_{d} \rangle$ from their sites on the superlattice  
characterizes the resulting configuration (very small 
values of $\langle r_{d} \rangle$ correspond to nearly perfect 
superlattices while $\langle r_{d} \rangle \sim a$ indicates an 
essentially ``melted'' state with randomized impurities).
Curie temperatures for a broad range of $\langle r_{d} \rangle$ 
are shown in panel (c) of Fig.~\ref{Fig:fig3}
(our superlattice is simple cubic 
with lattice constant $2a$, which corresponds to 
$x$ = 0.031).   
While the calculated $T_{c}$ in (c) slightly increases
as the superlattice is disrupted,
the effect is relatively modest, consistent with previous work 
comparing random disorder and pure superlattice 
configurations~\cite{docepuntocinco,docepuntosiete}.
The results suggest that while an excessive degree of clustering can lower
$T_{c}$, the absence of any inhomogeneities in the impurity configurations 
is also deleterious to the ferromagnetic state.  
As the transition is made from a perfect superlattice
to more disordered configurations, more and more impurities 
are displaced and can act as a bridge 
between neighboring superlattice sites, thereby increasing the effective 
coupling between neighboring Mn ions still occupying superlattice sites.  
Ultimately, the $T_{c}$ curve saturates for large $\langle r_{d} \rangle$
where superlattice modulations vanish.
It thus appears that the uncorrelated impurity distributions occupy a ``happy'' 
medium with a critical temperature higher than that of either strongly clustered
configurations or states with superlattice structure.  Thus random disorder 
modestly enhances DMS $T_{c}$.  

\section{The Strong Disorder Limit}
\begin{figure}
\centerline{\psfig{figure=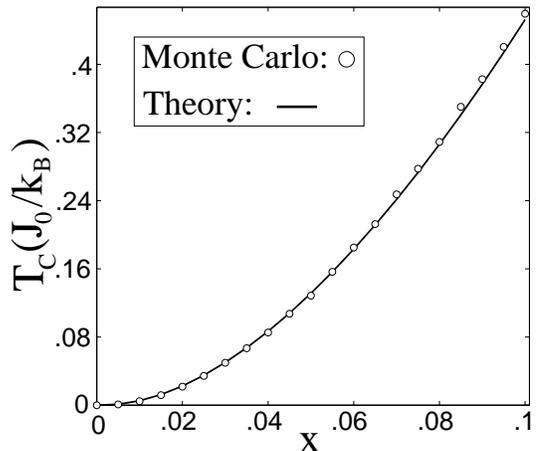,width=2.75in}}
\caption{The graph shows together the numerical MC $T_{c}$ results
(open circles) and the the theoretical $T^{*}$ (solid line) for
$n_{c}/n_{i}=0.1$ and $l/a=0.5$}
\label{Fig:fig4}
\end{figure}

\begin{figure}
\centerline{\psfig{figure=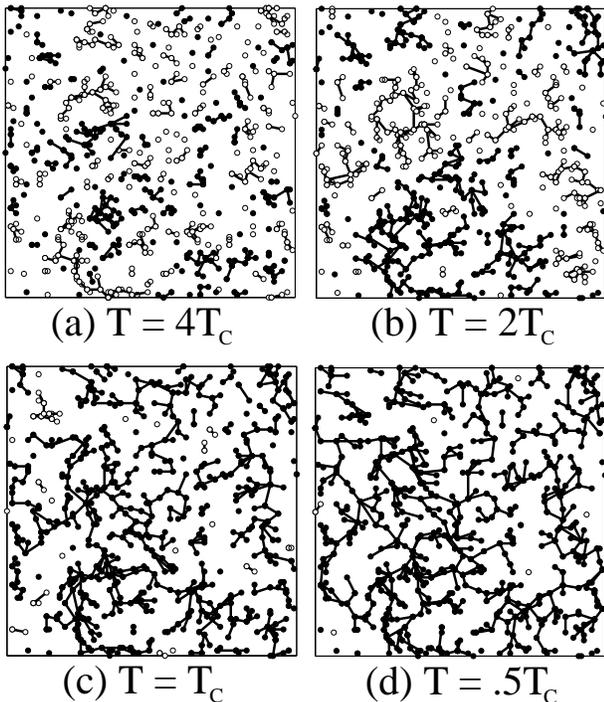,width=3.2in}}
\caption{The four panels (a,b,c, and d) display clusters
of correlated spins for $x = 0.05$, $J(r) = J_{0}e^{-r/l}$ where 
$l/a = 1.0$.  Each panel corresponds to a different temperature
with $T/T_{c} = \left \{4.0,2.0,1.0,0.5 \right \}$ for (a), (b), (c), 
and (d) respectively.  Dark/open circles represent ``up''/``down'' 
spins.}
\label{Fig:fig6}
\end{figure}

    Finally, we examine the regime of very strong disorder (i.e. 
$l_{s} \gg \left \{ l,a \right \}$), finding that
the Curie temperature exhibits well-defined asymptotic behavior in the 
large $l_{s}$ limit corresponding to the percolation of randomly 
placed spheres.  We first exploit the 
strong similarity of the behavior of discrete and continuum systems 
shown in Figure~\ref{Fig:fig2} by relaxing the discrete
occupancy requirement, although this condition is still imposed in the 
numerical calculations.  For convenience, we rescale the linear 
dimensions in such a way that $l_{s}$ is equal to unity, thereby 
mapping our problem to a new ferromagnetic system where the 
range of $J(r^{'})$ is shortened to $l^{'} = l/l_{s}$.  Hence, for 
large $l_{s}$, $J(r^{'})$ varies quite rapidly and as a result the 
coupling between pairs of spins tends either to be very large or very small.
The strong disorder limit thus can be viewed as the regime in which 
spins even slightly closer than the ``correlation radius'' $r_{\mathrm{corr}}$
are strongly correlated whereas for separations greater than $r_{\mathrm{corr}}$, the 
coupling rapidly becomes small relative to $k_{B}T$.
This is essentially a real space renormalization group (RG) argument for the 
scaling of the spin coupling.
This observation motivates a comparison with the problem of randomly 
placed spheres (distributed with unit density $n = 1$),
which form percolating networks when their radii exceed $r_{c}$; it 
is natural to identify the critical value of $r_{\mathrm{corr}}$ with $2r_{c}$ and to 
write $k_{B}T^{*} = J(r_{\mathrm{corr}})$ for the ferromagnetic transition
temperature~\cite{trece}.
However, a constant $\eta$ of order unity is needed to take thermal 
fluctuations into account~\cite{catorce}  and one actually has 
$T_{c} = \frac{\eta}{k_{B}}J(2r_{c}l_{s})$,
where the final form is expressed in terms of the original, unmodified 
units.  Our MC simulations give essentially exact (within $1-2\%$) quantitative 
agreement (shown in Fig.~\ref{Fig:fig4} for
the $x=0-0.1$ range we checked) between the numerical MC 
$T_{c}$ results and the theoretical $T^{*}$ with both showing essentially 
singular $x^{4/3}\exp(-\alpha x^{-1/3})$ 
behavior for small $x$.   
    We show explicitly in Fig.~\ref{Fig:fig6} that the approach to $T_{c}$ from above is 
marked by the formation of magnetic clusters of correlated spins by visualizing 
these clusters using the Swendsen-Wang algorithm~\cite{quince}
(which partitions the entire system into spin clusters).
To facilitate viewing, we examine a 2D 
system, a dilute site-disordered Ising model,
where ($x=0.05$ and $l/a = 1.0$).  In Fig.~\ref{Fig:fig6}, Swendson-Wang snapshots are 
depicted for four temperatures.  One can see the merging of disconnected
clusters to form a spanning cluster at $T_{c}$, though substantial
clustering occurs even well above $T_{c}$.  This is very similar
to the magnetic polaron percolation picture of DMS ferromagnetism 
theoretically envisioned in Ref.~\cite{menosseis} and the qualitative
discussion in Refs.~\cite{docepuntocinco,docepuntosiete}.  
We point out, however,
that the $T_{c}$ formula in our RG argument (validated by our MC simulations)
differs somewhat from the corresponding percolation theory result~\cite{menosseis}.
We emphasize that although our discussion has been in the DMS context, our 
arguments are not dependent on particular traits of the RKKY range 
function. 

\section{Conclusions}
    In conclusion, we have examined the effect of impurity correlation
in diluted magnetic semiconductors.
We find that the presence of correlations among impurity positions
affects the critical temperature $T_{c}$ only mildly, 
with the calculated $T_{c}$ being optimal for neutral impurity 
correlations (i.e. random disorder).
In addition, we have verified via explicit numerical calculation that 
just as ferromagnetic behavior in the weak disorder mean field limit is 
amenable to mean field approaches, the strong disorder regime 
also can be understood in terms of a limiting theory, random 
sphere percolation.  We develop a new theory for the dependence of 
$T_{c}$ on the magnetic impurity density by mapping the discrete 
random problem to a continuum problem using real space RG 
arguments, and validate the theory by Monte Carlo simulations.
This work is supported by US-ONR.

\end{document}